\documentclass[twocolumn]{revtex4}
\usepackage{latexsym}
\usepackage{graphics,epsfig,graphicx}
\usepackage{epsfig}
\usepackage[german, english]{babel}
\usepackage[latin1]{inputenc}
\usepackage{color}
\usepackage{makeidx}

\begin{document}

\title{A diode laser stabilization scheme for $\mathrm{^{40}Ca^{+}}$ single ion spectroscopy}

\author{Felix Rohde, Marc Almendros, Carsten Schuck, Jan Huwer, Markus Hennrich and J\"urgen Eschner}

\affiliation{ICFO-Institut de Ciències Fotòniques, Mediterranean
Technology Park, E-08860 Castelldefels (Barcelona), Spain}

\date{\today}

\begin{abstract}
We present a scheme for stabilizing multiple lasers at wavelengths
between 795 and 866~nm to the same atomic reference line. A
reference laser at 852~nm is stabilized to the Cs $\mathrm{D_2}$
line using a Doppler-free frequency modulation technique. Through
transfer cavities, four lasers are stabilized to the relevant
atomic transitions in $\mathrm{^{40}Ca^{+}}$. The rms linewidth of
a transfer-locked laser is measured to be 123~kHz with respect to
an independent atomic reference, the Rb $\mathrm{D_1}$ line. This
stability is confirmed by the comparison of an excitation spectrum
of a single $\mathrm{^{40}Ca^{+}}$ ion to an eight-level Bloch
equation model. The measured Allan variance of $10^{-22}$ at 10~s
demonstrates a high degree of stability for time scales up to
100~s.
\end{abstract}

\maketitle
\section{Introduction}
\label{intro} Since the first preparation of a single trapped
$\mathrm{Ba^+}$ ion \cite{Neuhauser1980PRAv22p1137a}, ion-trap
experiments have become increasingly sophisticated. Ultra high
precision spectroscopy has lead to single-ion frequency standards
\cite{Diddams2001Sv293p825,Schmidt2005Sv309p749} and today the
most accurate clock is based on single-ion spectroscopy
\cite{Rosenband2008Sv319p1808}. Trapped ions have also developed
into very promising candidates for the implementation of schemes
of quantum computation, and universal gate operations
\cite{Leibfried2003Nv422p412,Schmidt-Kaler2003Nv422p408,Home2006NJPv8p188}
as well as quantum algorithms
\cite{Chiaverini2004Nv432p602,Reichle2006Nv443p838,Brickman2005PRAv72p50306}
have been demonstrated. In quantum communication, first building
blocks of a quantum network have also been realized with ions
\cite{Olmschenk2009Sv323p486}. Besides these applications, ions
have proven to be textbook-like model systems to study fundamental
questions in quantum optics and quantum mechanics
\cite{Jost2009Nv459p683}.

In such experiments the ions are optically cooled and manipulated
by laser light. Depending on the chosen ion and the desired
application, the lasers have to be frequency-stabilized well below
the typical frequencies involved such as the natural linewidth
($\approx 20\,\mathrm{MHz}$) of the optical transition that is to
be driven, Zeeman splitting ($\approx 10\,\mathrm{MHz}$), or the
vibrational frequency in the trap ($\approx 1\,\mathrm{MHz}$).
Furthermore, data are typically collected over long periods of
time, such as hours or even days, since experiments exhibit low
count rates. A laser system used for single ion spectroscopy
therefore faces the requirements of offering good frequency
stability over both short and long periods of time.

In early experiments the choice of ion species was governed by the
availability of laser sources, typically dye lasers, at the
characteristic wavelengths of the ion. Technical progress and the
availability of a broad spectrum of wavelengths has made robust
and inexpensive diode lasers increasingly attractive. However,
significant technical effort to frequency-stabilize these lasers
has remained indispensable. A popular approach is to lock a laser
to a passive cavity built from ultra-low expansion material and
placed in a pressure-sealed container or in vacuum
\cite{C.Raab1998APBv67p683}. This eliminates the extreme
sensitivity of the cavity resonance to pressure changes due to the
dependence of the refractive index of air
\cite{Riedle1994RoSIv65p42}. Nevertheless such cavities are not
totally drift free, they only minimize the drift of the laser
frequency mostly caused by temperature fluctuations in the
environment. Other approaches use scanning transfer cavities in
combination with a stable HeNe reference laser. This technique is
slow compared to others as it is limited by the scanning frequency
of the cavity \cite{Matsubara2005JJoAPv44p229}. An open cavity
transfer scheme combining active piezo and temperature
stabilization of the cavity length has been used to build a
difference frequency spectrometer \cite{Riedle1994RoSIv65p42},
whereby a custom-made pyrex cavity transfers the stability of a
HeNe laser onto another laser with a cavity modulation technique.

A scheme similar to the one we will present here has been used for
high-precision frequency measurements of the $\mathrm{D_1}$ line
of alkali atoms \cite{A.Banerjee2004ELv65p172}: a reference laser
is locked to the Rb $\mathrm{D_2}$ line by saturated absorption
spectroscopy. Its 10~kHz linewidth is transferred to a second
laser of the same type via an in-vacuum ring cavity. This second
laser was then used to measure the absolute frequency of the Rb
$\mathrm{D_1}$ line using a second saturated absorption
spectroscopy set-up. The ultimate but also cost-intensive solution
for stable referencing is the use of recently developed frequency
combs. In ultra high precision spectroscopy the use of frequency
combs is becoming more and more common
\cite{Maric2008PRAv77p32502}.

In this article we present a scheme for the stabilization of four
diode lasers that are resonant with various transitions in
$\mathrm{^{40}Ca^{+}}$. All four lasers are stabilized with
respect to the $\mathrm{D_2}$ crossover line (F=3 $\rightarrow
\mathrm{F'=3/4}$) of atomic Cs using a Pound-Drever-Hall (PDH)
transfer lock
\cite{pound490,Drever1983APBv31p97,G.C.Bjorklund1983APBv32p145}.
The scheme employs a reference laser at the Cs wavelength and open
low-finesse transfer cavities built from inexpensive commercially
available components that avoid the need of a vacuum container and
ultra low expansion materials. A combined temperature and piezo
length stabilization of the cavities is controlled by self-built
locking electronics. In general this method can be readily
transferred to any desired wavelength by an appropriate coating of
the cavity mirrors.

\section{Stabilization scheme}
\label{Stabilization scheme} 
\begin{figure}
\begin{center}
\resizebox{0.7\columnwidth}{!}{
\includegraphics{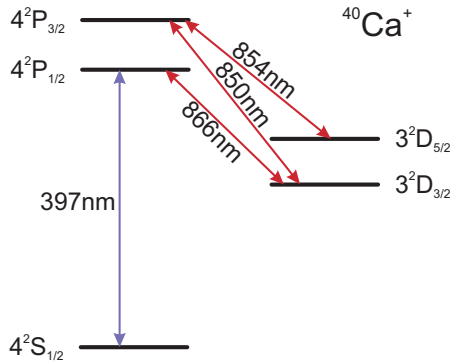}
 }\caption{Level scheme of $\mathrm{^{40}Ca^+}$. Diode lasers for the
 four indicated transitions are stabilized with the transfer lock.
 Light at 397~nm is produced by frequency doubling of a laser at 795~nm.}
\label{Termschema_Ca2}
\end{center}
\end{figure}

The level scheme of
$\mathrm{^{40}Ca^{+}}$ and the four relevant transitions are
displayed in figure \ref{Termschema_Ca2}.
All lasers are commercial grating-stabilized diode lasers. The
laser at 397nm is a frequency-doubled grating-stabilized diode
laser. The four lasers are stabilized using the same stabilization
scheme that consists of a series of consecutive PDH locks.

In this scheme, a reference laser at 852~nm is locked to the Cs D2
line using frequency modulation at 20~MHz in combination with
Doppler-free absorption spectroscopy in a Cs vapor cell. The
stability of this laser is transferred to the other lasers by the
use of open length-stabilized low-finesse cavities (section
\ref{cavities}). Each transfer cavity is brought into resonance
with both the reference laser and one of the
$\mathrm{^{40}Ca^{+}}$ lasers at the same time by temperature
tuning. A PDH lock stabilizes the length of the cavity with
respect to the reference laser. Each $\mathrm{^{40}Ca^{+}}$ laser
is then stabilized with another PDH lock to its respective
transfer cavity. The advantage of this scheme is that not only the
linewidth on a short time scale is reduced due to the lock to the
cavity, but also long term drifts are eliminated due to the
referencing to an atomic line.
\begin{figure}
\resizebox{\columnwidth}{!}{%
\includegraphics{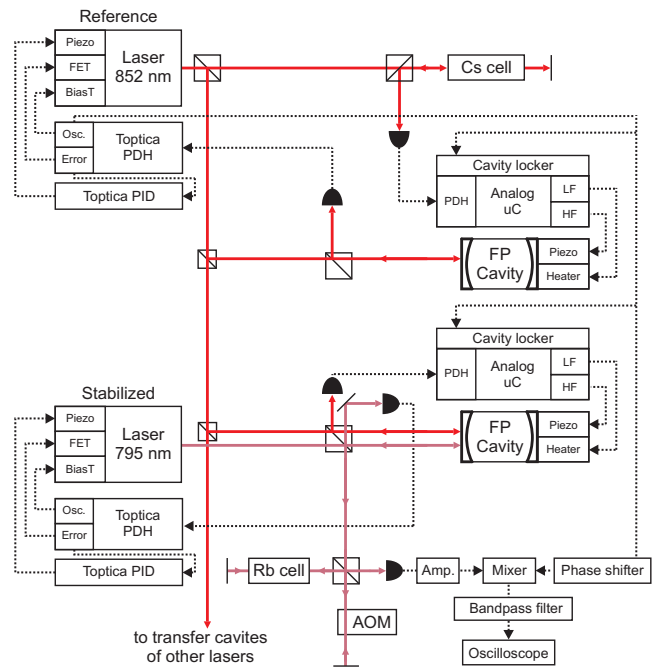}
 }\caption{Laser stabilization scheme for the 795~nm laser. The
 stability of a reference laser at 852~nm is transferred onto
 the 795~nm laser using transfer cavities. All relevant elements
 of the scheme are described in detail in the text.}
 \label{StabSchemeRb}
\end{figure}

As an example, figure \ref{StabSchemeRb} shows a schematic of the
locking chain for the 795~nm laser, the master laser for the
397~nm system. A beam splitter sends part of the 852~nm reference
beam to a Cs vapor cell and the remaining power to the transfer
cavities. The error signal derived from the saturated absorption
spectroscopy on the Cs cell is fed back onto a piezo controlling
the length of the first cavity via a self-built electronic device
for stabilization called "cavity locker" (section \ref{Cavity
Locker}). The cavity locker has a PDH module that uses the 20~MHz
oscillation signal from a current modulation of the reference
laser to demodulate the absorption signal. A digitally implemented
servo amplifier controls the piezo of the cavity. At the same time
a low frequency regulation stabilizes the temperature via a
heating wire. The error signal from the light reflected from this
first cavity is used to stabilize the reference laser to this
cavity with the help of commercial electronics.

The transfer cavity of the 795~nm laser is stabilized to the
reference laser employing the error signal derived from the
reflected 852~nm beam. A second cavity locker, again using the
local oscillator of the reference laser for demodulation, feeds
the signal back onto heater and piezo of the cavity. The 795~nm
laser itself is stabilized to the transfer cavity via its
reflection signal. The three other lasers are stabilized, by means
of additional transfer cavities and cavity lockers, in exactly the
same way.

In the following, we describe the relevant elements of the
stabilization scheme in detail.

\section{Cavities}
\label{cavities} The confocal Fabry-Perot cavities are built from
two mirrors \footnote{Layertec} ($\mathrm{T=99,5 \%, radius=30
cm}$) with 15~cm distance and have 500~MHz free spectral range.
Finesse and linewidth are measured to be 270 and 1.9~MHz,
respectively.

\begin{figure}
\resizebox{\columnwidth}{!}{%
\includegraphics{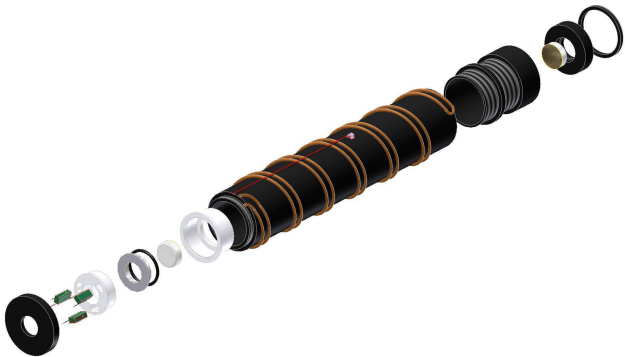}
} \caption{Cavity assembly consisting of a commercial lens tube
system and a custom-made teflon holder. The picture shows from
left to right: end cap with hole for optical access and electric
connections, three piezo stacks, teflon piece guiding the piezos,
aluminium washer, rubber ring, first mirror, teflon piece holding
the aluminium washer and rubber ring, aluminium tube with heating
wire and temperature sensor, end piece for course adjustment,
second mirror, mirror holder, threaded ring to fix the mirror
holder in the end piece.} \label{Ensemble-cavity}
\end{figure}

The two mirrors are mounted at the two ends of an aluminium lens
tube \footnote{Thorlabs} as depicted in Figure
\ref{Ensemble-cavity}.
Having a high temperature expansion coefficient, aluminium assures
the temperature tunability of the mirror distance
($\mathrm{\frac{\Delta L}{\Delta T}=3.5}\,\mathrm{\mu
m/^{\circ}C}$). A heating wire with $15\,\mathrm{\Omega}$
resistance is wound around the tube and a PT100 temperature sensor
is attached. The whole tube itself is wrapped in insulating
material and placed inside a bigger aluminium container that is
mounted on the optical table. One mirror is glued to an end piece
of the tube that can be screwed in and out for course length
adjustment. The other mirror is glued to an aluminium washer that
is mounted in a two-piece teflon holder together with 3 piezo
actuators and a rubber ring as shown in Figure
\ref{Ensemble-cavity}. The holder is designed such that the
aluminum washer with the mirror is pressed by the piezo stacks
\footnote{Piezomechanik PSt 150/2x3/7} against the rubber ring
\footnote{inner diameter = 16~mm, thickness = 1.5~mm} that sits in
between the aluminum washer and the outer teflon part. The piezo
stacks are guided by three holes in the inner teflon part that
sits inside the outer one. They are held by an end cap that is
screwed onto the aluminium tube and pre-loads the flexible rubber
ring. Thereby the piezos find sufficient resistance to compress
the rubber ring thus changing the length of the cavity. A maximum
voltage of $\pm 10\,\mathrm{V}$ is applied to the piezos
corresponding to a length change of about $2.5\,\mathrm{\mu m}$
and resulting in a shift of the resonance over about three free
spectral ranges. Scanning over a wider frequency range is attained
by changing the temperature.

\section{Cavity Locker}
\label{Cavity Locker}
\begin{figure}
\begin{center}
\resizebox{0.7\columnwidth}{!}{%
\includegraphics{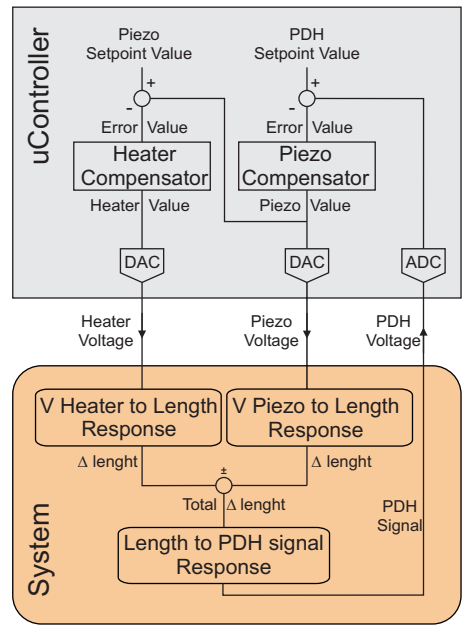}
} \caption{Block diagram of the control loop used to stabilize the
cavities. The system consists of the cavity with its piezos and
heater that control the length and the resulting PDH signal
generated by the PDH module. The micro controller processes the
error signal and controls the piezos and the heater. ADC and DAC
stand for analog-to-digital and digital-to-analog converter. The
Pound-Drever-Hall (PDH) setpoint value is programmed to coincide
with the zero crossing of the error signal. The piezo setpoint
value is adjusted to the piezo mid position (0~V) and the zero
crossing. If the control loop is closed, the heater compensator
reacts according to the piezo signal and the piezo setpoint.}
\label{CavityLockerCLoopAI}
\end{center}
\end{figure}

The stabilization of the cavities to the reference laser is
realized with a self-built electronic device called "cavity
locker" \cite{Almendros2009}. The cavity locker consists of a PDH
input stage, a microcontroller\footnote{Analog Devices
EVAL-ADUC7026QSZ}, a scanning unit, and outputs for low- and
high-frequency feedback (figure \ref{StabSchemeRb}). The PDH stage
consists of an input amplifier, a mixer, a phase shifter, and a
bandpass filter, and it demodulates the photo diode signal using
the rf modulation signal of the reference laser current as local
oscillator (figure \ref{StabSchemeRb}). The PDH error signal which
it produces is the input for the servo amplifier which is
implemented with software on the microcontroller. In control mode
the low frequency output of the servo amplifier drives the heating
wire, and the high frequency output drives the cavity piezos. If
the control loop is open, the piezo can be scanned by the scanning
unit and the temperature is stabilized to the value defined by a
temperature set point potentiometer.
\begin{figure*}
\begin{center}
\resizebox{1.9\columnwidth}{!}{%
\includegraphics[angle=-90]{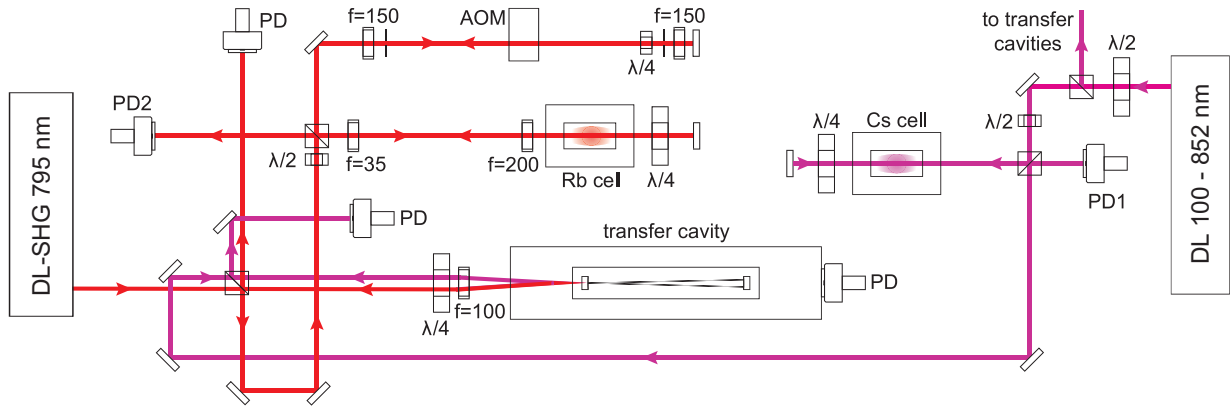}
} \caption{Setup for the characterization of the transfer lock.
The master laser of a frequency doubling stage at 795~nm is locked
to a Doppler-free Cs resonance in a gas cell by three consecutive
Pound-Drever-Hall stabilizations using a transfer cavity. To
characterize the frequency stability of the laser a PDH error
signal is derived from Doppler-free absorption spectroscopy on
$\mathrm{^{85}Rb}$.} \label{Setup Rb spectroscopy}
\end{center}
\end{figure*}

The servo amplifier implemented on the evaluation board consists
of two compensators, one for the piezos and one for the heating
wire. In closed-loop operation the heater compensator is
programmed to regulate the length of the cavity in such a way that
the piezos remain in their mid position (figure
\ref{CavityLockerCLoopAI}). This assures that the piezo voltage is
kept within $\pm 10\,V$. Both compensators are programmed (in
$\mathrm{C}$ language) after analyzing the response of the system.
The speed of the piezo compensator loop is limited by the
eigen-frequency of the piezo actuators measured to be 3~kHz.

\section{Characterization}
\label{Characterization}
\subsection{Cs spectroscopy}
\begin{figure}[b]
\begin{center}
\resizebox{\columnwidth}{!}{%
\includegraphics{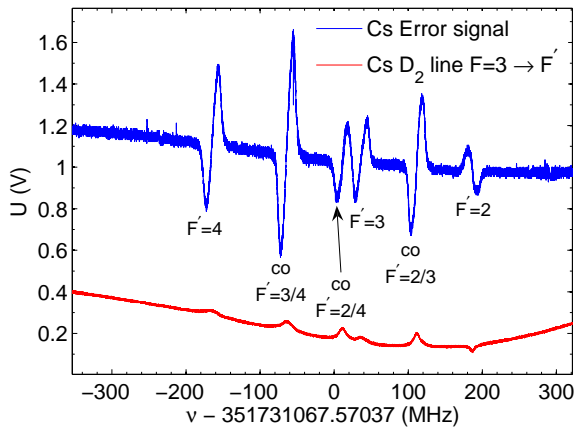}
} \caption{Cs Doppler-free absorption spectroscopy signal (bottom)
and corresponding error signal (top) for the $\mathrm{F=3
\rightarrow F'=2,3,4}$ transitions and its cross-over lines. The
abscissa shows absolute frequency. }
\label{CsAbsorpErrorSignal_gauge}
\end{center}
\end{figure}

The right part of figure \ref{Setup Rb spectroscopy} shows the
set-up of the Doppler-free absorption spectroscopy on the Cs cell.
The absorption signal detected with PD1 is shown in figure
\ref{CsAbsorpErrorSignal_gauge} together with the corresponding
error signal produced by the Pound-Drever-Hall stage of the cavity
locker.
The transition used as reference is the $\mathrm{F=3}$
$\rightarrow$ $\mathrm{F^\prime=3/4}$ crossover line.

\subsection{Individual stabilizations}
\label{Individual stabilizations} To characterize the stability of
the individual Pound-Drever-Hall locks of the lasers to the
cavities, the root mean square frequency deviation with respect to
the cavities are measured. For this purpose the error signal of
each laser was fitted with the corresponding theoretical curve and
the slope at the set point was extracted; the frequency axis was
gauged via the sidebands arising from the modulation of the laser
current. A root mean square frequency deviation of 38~kHz was
obtained, measured from the in-loop error signal recorded over
2~ms.

\subsection{Characterization of the transfer lock}
\label{Characterization_transfer} To estimate the absolute
stability of the transfer locked lasers, one of them was compared
with an independent atomic reference, a Doppler-free rubidium
resonance in a gas cell. For this purpose a second Doppler-free
absorption spectroscopy was set up and the master laser of the
397~nm SHG system was tuned to the D1 line of $^{85}\,\mathrm{Rb}$
at 794.979~nm \cite{Banerjee2003OLv28p1579}. In figure \ref{Setup
Rb spectroscopy} this setup is shown.

\begin{figure}
\begin{center}
\resizebox{\columnwidth}{!}{%
\includegraphics{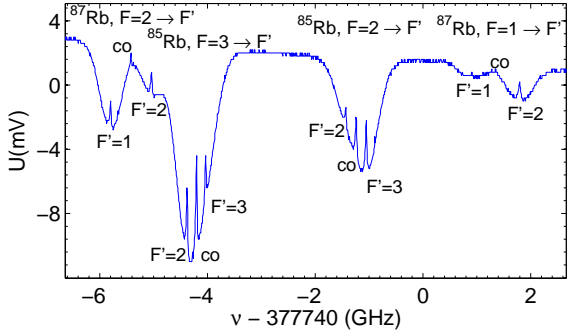}
} \caption{Full spectrum of the Rb $\mathrm{D_1}$ line, obtained
via Doppler-free absorption spectroscopy. }
\label{fullspec_rb_avFreq}
\end{center}
\end{figure}

The full Rb Doppler-free absorption signal obtained by scanning
the frequency of the 795~nm laser is depicted in figure
\ref{fullspec_rb_avFreq}. Since the cell contains Rb isotopes in
their natural abundances, absorption lines for $^{85}\mathrm{Rb}$
(72.2~\%) and $^{87}\mathrm{Rb}$ (27.8~\%) are observed. The
Doppler-free hyperfine structure is well resolved. The frequency
axis was gauged using the literature value \cite{Steckvp} of the
$\mathrm{5^2S_{1/2}}$ ground state hyperfine splitting of
$\mathrm{^{87}Rb}$ of $\delta\nu=6.834682 \,\mathrm{GHz}$. The
different hyperfine dips are easily identified using the
splittings known from the literature \cite{Steckvp,Steckvpa}.

The transfer lock is characterized by generating a PDH error
signal from the saturated absorption spectroscopy signal on
$^{85}\mathrm{Rb}$, using the modulation of the laser light at
20~MHz from the lock to its transfer cavity. The signal recorded
with a fast photo diode (PD2) \footnote{New Focus model 1801
(Bandwidth 125~MHz)} behind the Rb cell is amplified, mixed down
with a mixer \footnote{Mini Circuits ZAD-1H} and filtered by a
bandpass filter \footnote{Mini Circuits SIF-21.4+} (see figure
\ref{StabSchemeRb} bottom).

\begin{figure}
\begin{center}
\resizebox{\columnwidth}{!}{%
\includegraphics{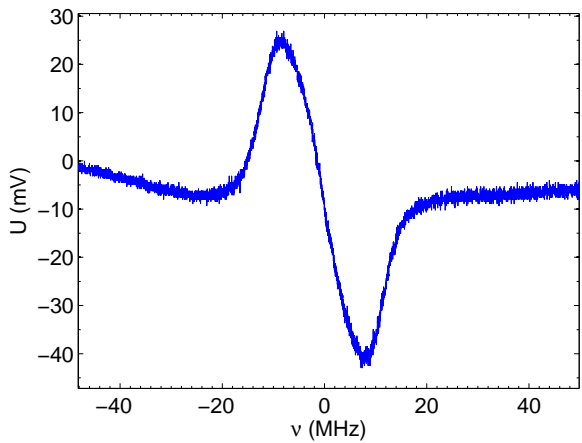}
} \caption{Error signal of the $\mathrm{^{85}Rb, F=3 \rightarrow
F'=2/3}$ crossover line. The signal was recorded by scanning an
AOM with the laser locked to an atomic Cs reference. }
\label{Rbfringe}
\end{center}
\end{figure}

With the whole stabilization chain locked, i.e. the master laser
at 795 nm is locked via the transfer cavity to the stable
reference laser, the laser was tuned to resonance with the
strongest line in $\mathrm{^{85}Rb}$, the $\mathrm{F=3}$
$\rightarrow$ $\mathrm{F'=2/3}$ crossover line. Scanning the AOM
of figure \ref{Setup Rb spectroscopy} we identified the desired
transition and set the frequency to resonance, i.e. to the point
with the steepest slope of the error signal. Figure \ref{Rbfringe}
shows the error signal of the $\mathrm{F=3 \rightarrow F'=2/3}$
crossover line obtained by scanning the AOM.

The error signal also serves as a frequency gauge for converting
the voltage recorded with the oscilloscope into frequency. For
this purpose the AOM is linearly scanned with $\pm
2\,\mathrm{MHz}$ amplitude around the resonance. The recorded
slope is fitted linearly to determine the conversion factor
between frequency and voltage for each measurement.

To measure the short-term stability of the locking chain, the Rb
error signal was recorded for a fixed AOM frequency at three
different times during one day over a time span of 200~ms with
different sampling rates. From the voltage trace on the
oscilloscope, rms deviations of 143~kHz, 123~kHz and 133~kHz were
measured, respectively. Recalling the result of 38~kHz for a
single PDH lock from section \ref{Individual stabilizations}, the
stability of 123~kHz for three consecutive locks is consistent
with the former measurement. However, the result from the
spectroscopy measurement is more relevant as it has been measured
with respect to an independent reference and can thus be assumed
to hold for the $\mathrm{Ca^+}$ spectroscopy.

In order to measure the long term stability of the setup, the mean
value of the voltage of the Rb error signal was recorded with a
computer. A Labview program read out the voltage from the
oscilloscope and stored the mean value over the time $\Delta
\tau=1\,\mathrm{s}$ for a duration of 2 hours. The mean value
($\Delta \tau=1\,\mathrm{s}$) of the error signal drifted by about
one short-term rms deviation ($\Delta t=200\,\mathrm{ms}$), i.e.
about 130~kHz, during these two hours.

The main cause of instability are acoustic noise or pressure
drifts, to which open transfer cavities are particularly
sensitive: there is a remaining systematic error resulting from
the different refractive indices of air at the wavelengths of the
two lasers \cite{Riedle1994RoSIv65p42}. Since among the
$\mathrm{Ca^+}$ lasers and the Cs reference the difference in
refractive indices is largest for 852 and 795~nm, the error for
the 795~nm laser can be regarded as an upper limit for all lasers.

An analysis following Ref. \cite{Riedle1994RoSIv65p42} reveals
that the observed drift is equivalent to a pressure change of
1~mbar. Since the pressure in our lab is not controlled, it is
reasonable to assume that the frequency drift is caused by a
pressure drift. Apart from dynamic fluctuations, the daily cyclic
variation of the atmospheric pressure is on the order of 1~mbar
for our latitudes. This sensitivity to pressure fluctuation is the
main limitation of schemes using open transfer cavities.

\subsection{Allan variance}
\begin{figure}
\resizebox{\columnwidth}{!}{%
 \includegraphics{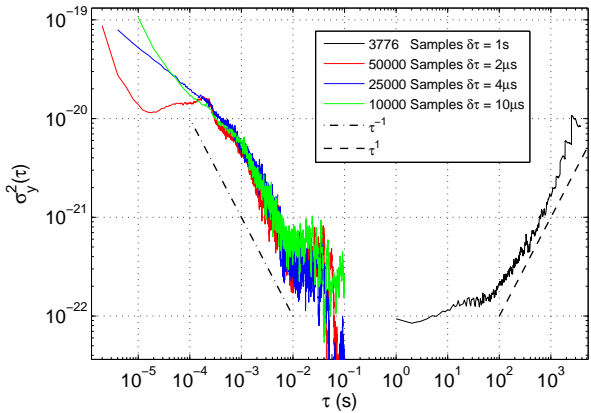}
} \caption{Allan variance for three short-term measurements with
different sampling rate and 200~ms integration time, and one
long-term measurement with 2h integration time. The dashed-dotted
line ($\tau^{-1}$) indicates the dominating white frequency noise
for high frequencies. The dashed line ($\tau^{1}$) indicates the
dominating random walk frequency noise for long time scales. The
sharp drop of two of the short-term measurements for time values
close to $\tau=10^{-1}\,\mathrm{s}$ is purely statistical due to
the low number of points.} \label{Allanvariance_all_release}
\end{figure}

For a full characterization of the stability of the 795nm laser
oscillator, the Allan variance \cite{Allan1966} was calculated
from the three short term measurements as well as from the long
term measurement. The four results are plotted in figure
\ref{Allanvariance_all_release}. The three measurements over
200~ms show good agreement between $\tau=10^{-4}$ and
$\tau=10^{-2}\,\mathrm{s}$. The Allan variance $\sigma^2_y$ decays
with $\tau^{-1}$ over these two orders of magnitude, down to the
$10^{-22}$ region. This indicates the presence of white frequency
noise caused in electronic components. It is expected that the
accuracy of an oscillator increases with further integration time
until other effects, such as long term drifts, start to dominate
\cite{Riehle2004}. In accordance with this the Allan variance for
the long term measurement rises again starting from
$\tau=10^2\,\mathrm{s}$ up to $\tau=10^4\,\mathrm{s}$. The slope
is proportional to $\tau^1$ indicating that the dominant noise
type is a random walk of frequency noise caused by the
environmental conditions. This is in agreement with the
observation of high sensitivity to pressure fluctuations.

Long and short-term measurements connect well for time scales of
$\tau=10^{-2}$ to $10^1\,\mathrm{s}$. Although no data was
recorded between 0.1 and 1~s, a flat behavior of $\sigma_y^2
\propto \tau^0$ may be assumed in that region. This suggests an
expected flicker noise floor for low frequencies.

For the long term measurement each data point was averaged by the
oscilloscope over 1~s, with no dead time between subsequent
points. This assures that no fast oscillations were disregarded by
the measurement apparatus. The three short term measurements were
only sampled, i.e. each data point was recorded after the sampling
period specified in figure \ref{Allanvariance_all_release}. To
make sure that there is no high frequency noise present, an
additional Fast Fourier Transform (FFT) of the Rb error signal was
recorded with the oscilloscope.
\begin{figure}
\begin{center}
\resizebox{\columnwidth}{!}{%
\includegraphics{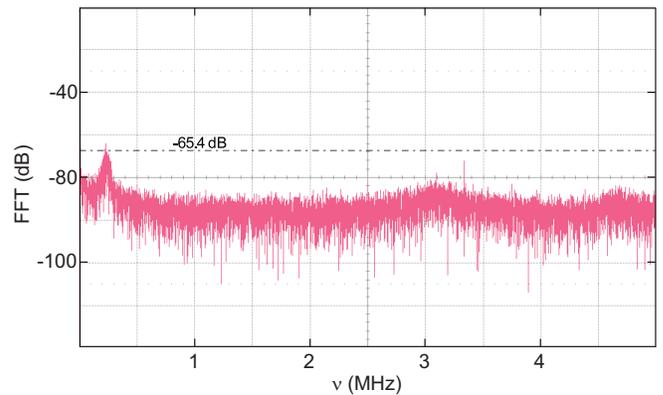}
} \caption{Fast Fourier Transform (FFT) of the Rb error signal for
a bandwidth of 5~MHz.} \label{FFT_rb_error2_5mhz}
\end{center}
\end{figure}
Figure \ref{FFT_rb_error2_5mhz} shows the FFT for a bandwith of 5
MHz. All noise with higher frequency than that is filtered by the
bandpass filter (figure \ref{StabSchemeRb}). The most prominent
peak at 250~kHz with an amplitude of -65.4~dB is still accounted
for in the Allan variance measurement with sampling rate
$2\,\mathrm{\mu s}$. For higher frequencies there is no
significant noise present. Therefore sampling without averaging on
the time scale of the short-term measurements is expected to not
miss any higher-frequency noise contributions.

\section{Single ion spectroscopy}
\begin{figure}
\begin{center}
\includegraphics[width=\columnwidth]{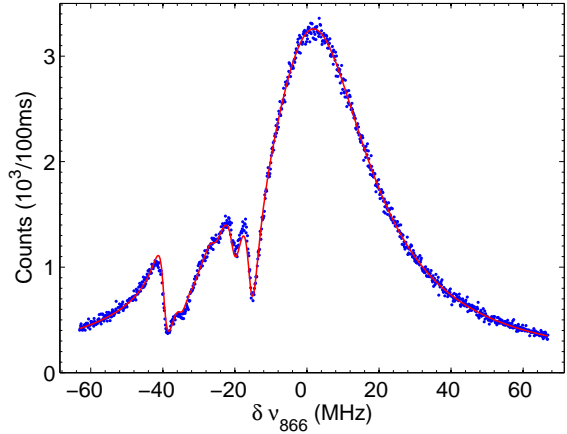}
\caption{Excitation spectrum of a single $\mathrm{^{40}Ca^+}$ ion
in a linear Paul trap with excitation under $45^{\circ}$ to the B
field and fluorescence detection in the direction of the B field.
The (red) solid line is an eight level Bloch equation model. From
the fitted curve the experimental parameters are calibrated as
follows: Rabi frequencies
$\mathrm{\Omega_{397}=2\pi\cdot15\,MHz}$,
$\mathrm{\Omega_{866}=2\pi\cdot3.5\,MHz}$, detuning
$\mathrm{\Delta_{397}/2\pi\cdot=-27\,MHz}$ and magnetic field
$\mathrm{B=3.8\,G}$. The fit also accounts for slight deviations
of the laser polarization from their ideal values ($< 3\,\%$).}
\label{fit200901161722}
\end{center}
\end{figure}

To test the stabilization scheme, excitation spectra of single
$\mathrm{^{40}Ca^+}$ ions have been recorded using the ion trap
set-up described in \cite{Gerber2009NJoPv11p13}. The ion is
continuously excited by two lasers at $397$~nm and $866$~nm, and
the fluorescence light of the ion is collected by a high numerical
aperture lens also described in \cite{Gerber2009NJoPv11p13}, and
detected with a PMT while the 866~nm laser is scanned over the
resonance. The $397\,\mathrm{nm}$ laser is red-detuned to provide
Doppler cooling. Figure \ref{fit200901161722} shows the excitation
spectrum of a single ion with excitation under $45^{\circ}$ to the
B-field direction. The $397$ and $866\,\mathrm{nm}$ beams were set
to vertical and horizontal polarization, respectively. Each data
point represents the count rate on the PMT integrated over
$100\,\mathrm{ms}$. The complex structure of the spectrum is
caused by two-photon (dark) resonances between the
$\mathrm{S_{1/2}}$ and the $\mathrm{D_{3/2}}$ level.

The red line is an eight-level Bloch equation model calculation
\cite{Schubert1995PRAv52p2994} that was fitted to the data with a
numeric (Matlab-based) fitting routine. The agreement with the
model is very good exhibiting a reduced $\chi^2$ of 1.8. The
linewidths of the lasers extracted from the fit are
$\delta\nu_{397}=268\,\mathrm{kHz}$ for the blue laser and
$\delta\nu_{866}=1/2 \cdot \delta\nu_{397} = 134\,\mathrm{kHz}$
for the infrared laser. This is consistent with the linewidth
measured with respect to Rb, as described in section
\ref{Characterization_transfer}.

\section{Conclusions}
\label{Conclusion} We have described a laser frequency
stabilization scheme for $\mathrm{^{40}Ca^+}$ single-ion
spectroscopy with high short- and long-term stability. The
residual rms deviation of 123~kHz is well below all significant
transition linewidths of $^{40}\mathrm{Ca^+}$ and allows us to
resolve well dark resonances in the excitation spectra. The Allan
variance proves high long-term stability limited only by pressure
changes in the laboratory. The scheme is particulary well suited
for measurements with low count rates that require stable
conditions over a long time period. For the longest experimental
run until the publication of this article, stable conditions for
40 hours were achieved. During this time no noticeable frequency
drift on the MHz scale was observed.

The linewidth of the laser system is comparable with systems used
for ultra-high precision spectroscopy
\cite{Maric2008PRAv77p32502}. The advantage in comparison with
other transfer stabilization schemes lies in the use of open
table-top cavities. The cavities are built mostly from
off-the-shelf parts, no vacuum or ultra low expansion material is
needed. A combined temperature and piezo length stabilization of
the transfer cavities controlled by a micro controller is at the
heart of the system.

We thank F. Dubin for technical support.

We acknowledge support from the European Commission (SCALA,
contract 015714; EMALI, MRTN-CT-2006-035369), the Spanish MICINN
(QOIT, CSD2006-00019; QLIQS, FIS2005-08257; QNLP, FIS2007-66944),
and the Generalitat de Catalunya (2005SGR00189; FI-AGAUR
fellowship of C.S.).

\end{document}